\documentclass[aps,prb,twocolumn,reprint,superscriptaddress,amsmath,amssymb,showpacs,citeautoscript]{revtex4-1}
\usepackage{stmaryrd}
\usepackage{amsmath}
\usepackage{amssymb}
\usepackage{graphicx}
\usepackage{dcolumn}
\usepackage{bm}
\usepackage{sidecap}

\begin{document}

\begin{titlepage}

\title{Spontaneous edge-defect formation and defect-induced conductance suppression in graphene nanoribbons}

\author{Jia Li}
\altaffiliation[Present address: ]{Fritz-Haber-Institut der
Max-Planck-Gesellschaft, Faradayweg 4-6, 14195 Berlin, Germany}
\author{Zuanyi Li}
\altaffiliation[Present address: ]{Department of Physics, University
of Illinois at Urbana-Champaign, Urbana, Illinois 61801, USA}
\author{Gang Zhou}
\affiliation{Department of Physics, Tsinghua University, Beijing
100084, People's Republic of China}
\author{Zhirong Liu}
\affiliation{College of Chemistry and Molecular Engineering, Peking
University, Beijing 100871, People's Republic of China}
\author{Jian Wu}
\author{Bing-Lin Gu}
\affiliation{Department of Physics, Tsinghua University, Beijing
100084, People's Republic of China}
\author{Jisoon Ihm}
\affiliation{Department of Physics and Astronomy, Seoul National
University, Seoul 151-747, Republic of Korea}
\author{Wenhui Duan}
\email[\*Corresponding author: ]{dwh@phys.tsinghua.edu.cn}
\affiliation{Department of Physics, Tsinghua University, Beijing
100084, People's Republic of China}

\date{\today}

\begin{abstract}

We present a first-principles study of the migration and
recombination of edge defects (carbon adatom and/or vacancy) and
their influence on electrical conductance in zigzag graphene
nanoribbons (ZGNRs). It is found that at room temperature, the
adatom is quite mobile while the vacancy is almost immobile along
the edge of ZGNRs. The recombination of an adatom-vacancy pair leads
to a pentagon-heptagon ring defect structure having a lower energy
than the perfect edge, implying that such an edge-defect can be
formed spontaneously. This edge defect can suppresses the
conductance of ZGNRs drastically, which provides some useful hints
for understanding the observed semiconducting behavior of the
fabricated narrow GNRs.

\end{abstract}


\maketitle

\end{titlepage}

\section{Introduction}

Graphene, a two-dimensional honeycomb lattice of carbon atoms, has
been regarded as a new promising material for fundamental studies
in condensed matter physics since its successful synthesis in
recent years \cite{review}. Subsequently, graphene nanoribbons
(GNRs), a quasi-one-dimensional structure of graphene with varying
widths, were fabricated by using several methods including e-beam
lithography \cite{Chen1,Han1}, chemical method \cite{Dai},
metallic nanoparticle etching \cite{Datta1}, and microscope
lithography \cite{Tapaszto,Weng}. As an electronic material, GNRs
were found to have some excellent properties such as simple
width-dependent band gaps, high carrier mobility at
room-temperature and so on, which make them suitable for device
applications in nanoelectronics
\cite{Chen1,Han1,Dai,Son,Barone1,Yan1,Beenakker,Zuanyi1}.

To realize practical applications of GNRs in electronics, their
widths must reach true nanometer scale ($\leq$10 nm), and
consequently, edge effects become crucial \cite{Tapaszto,Ritter}.
Whereas, current GNRs fabricated in experiments inevitably have edge
roughness, and their precise atomic edge structures are still
unclear and need to be further determined \cite{Dai,Datta1}.
Interestingly, recent experiments showed that all sub-10-nm GNRs are
semiconducting \cite{Dai} while previous theoretical works predicted
that GNRs with perfect zigzag edges are metallic \cite{edge-state}.
This discrepancy was suggested to originate from complicated edge
structures in practice. So far, a lot of theoretical efforts were
focused on the electronic structure of edge disorder by using the
tight-binding method \cite{disorder}. In contrast, the evolution and
thermodynamics of atomic-scale edge defects have been seldom
studied, especially via the first-principles approach. Therefore,
exploring possible stable edge defect structures and their influence
on the transport properties of GNRs is highly desired and helpful to
understand recent experimental observations of GNRs and make use of
GNRs in future nanoelectronics.

In this paper, with the density-functional theory (DFT)
calculations, we systematically investigate the migration of typical
edge defects (carbon adatom and vacancy) and the recombination of
the adatom-vacancy pair (AVP) in zigzag GNRs (ZGNRs). We find that
an adatom can migrate along the ribbon edge at room temperature
while a vacancy cannot. The migration of carbon adatom may lead to
the recombination of the AVP and the resulting pentagon-heptagon
ring (R57) defect structure is more stable than the perfect edge
structure. This localized R57 defect is more likely to exist at the
edges of experimentally fabricated ZGNRs than periodic R57 edge
structure recently proposed \cite{Wassmann1,Koskinen1}. We also find
that the presence of edge defects causes a decrease in the current
through the ZGNR under bias voltage. Especially, for ZGNRs with the
R57 edge defect, the current becomes very small and the conductance
approaches zero near the Fermi level. This may play an important
role in the observed semiconducting behavior of the fabricated
narrow GNRs (especially those with zigzag crystallographic edges)
\cite{Han1,Dai,Tapaszto}.

\section{method and model}

\begin{SCfigure*} 
\centering
\includegraphics[width=0.57\textwidth]{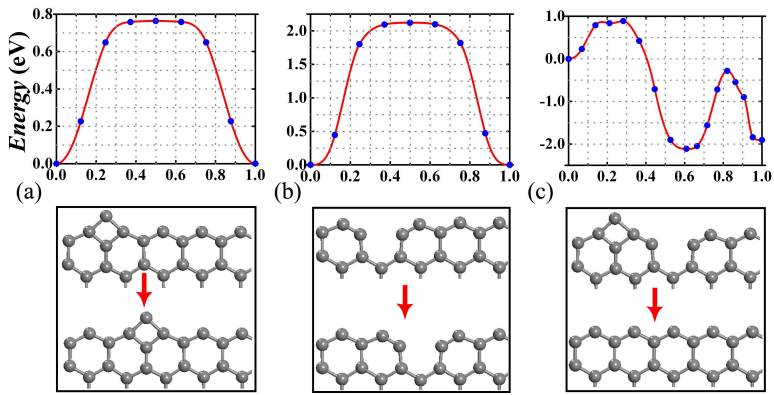}
\caption{\label{fig:fig1} (Color online) The MEP for (a) the carbon
adatom migration, (b) the vacancy migration and (c) the
recombination of AVP located at one edge of the 7-ZGNR. The $x$ axis
is the reaction coordinate in each case. The corresponding initial
and final structures are shown in the lower panel.}
\end{SCfigure*}

Our electronic structure calculations are performed using the Vienna
\emph{ab initio} simulation package (\textsc{vasp}) (Ref.~
\onlinecite{Kresse1}) within the framework of DFT. The projector
augmented wave potential \cite{Kresse2} and the generalized gradient
approximation with Perdew-Burk-Ernzerhof functional \cite{Perdew1}
are used to describe the core electrons and the exchange-correlation
energy, respectively. The cutoff energy for the plane wave basis set
is set to 400 eV. The energies are converged to below $10^{-5}$ eV
and the residual forces on all atoms are converged to below 0.01
eV/\AA. The minimum-energy path (MEP) calculations for the migration
of edge defects have been performed using climbing image nudged
elastic band method \cite{Henkelman1}. The quantum transport
calculations are performed using an \emph{ab initio} code,
\textsc{atomistix toolkit 2.0} (\textsc{atk 2.0}) \cite{Brandbyge1},
which implements non-equilibrium Green's-function formalism with
time-independent DFT \cite{DFT-NEGF-LIM}. This approach is limited
to the steady-state regime and based on a self-consistency
procedure. Such limitations makes the approach different from
Keldysh's non-equilibrium Green's function (see Ref.~
\onlinecite{Stefanucci}). Since the migration of edge defects are
dominated by the atoms nearby, in the calculation of the migration
path, we choose ZGNRs with the defect locating at one ribbon edge
and the other edge being perfect as the models. To exclude the
interaction between defects, we use a supercell including eight unit
cells of the ZGNR with the length along the ribbon axis of about
17.95 \AA. In the transport calculations, considering that every
defect has the same possibility of being located at the two edges in
experiments, we choose ZGNRs with both edges having one defect as
the models.

\section{Results and discussion}

We start by investigating the stable structures of adatom and
vacancy defects at the edge of ZGNRs. As a result of the $sp^2$
hybridization in a graphene sheet, each edge atom of a perfect ZGNR
has one dangling $sp^2$ orbital. When a carbon adatom is located at
the ZGNR edge, it prefers to stay at the bridge site between two
adjacent edge atoms to form two $\sigma$ bonds [the lower panel of
Fig.~\ref{fig:fig1}(a)]. However, when a vacancy is created at the
edge site of the ZGNR by removing an edge carbon atom, the two
carbon atoms adjacent to the vacancy, each having one dangling
$sp^2$ bond, move away a little along opposite directions and form
stronger bonds with their adjacent edge atoms [the lower panel of
Fig.~\ref{fig:fig1}(b)]. And the bond length decreases from 1.38 to
1.25 \AA. We further calculate the MEP for the migration of adatom
and vacancy defects along the edge of the 7-ZGNR (with seven zigzag
chains). The migration barrier ($E_{\rm B}$) obtained for the edge
adatom is 0.76 eV [the upper panel of Fig.~\ref{fig:fig1}(a)].
Although $\sigma$-bond breaking and rebonding are involved in this
process, the free space in the edge area allows for an easy
relaxation of neighboring atoms and substantially lowers the
barrier. This barrier is only 0.30 eV higher than the diffusion
barrier of a carbon adatom moving over a graphene surface
\cite{Lehtinen1}. On the other band, the calculated migration
barrier of the vacancy is about 2.12 eV [the upper part of
Fig.~\ref{fig:fig1}(b)], over three times of the barrier of carbon
adatom, suggesting that the edge vacancy is much less mobile than
the adatom.

In a thermodynamic sense, the rate coefficient $k$ for the migration
of edge defects is expressed as $\nu_{\rm G}\exp (-E_{\rm B}/k_{\rm
B}T)$, where the attempt frequency $\nu_{\rm G}$=${\prod_{i=1}^{3N}
\nu_{i}^{IS}}/{\prod_{i=1}^{3N-1} \nu_{i}^{TS}}$ can be calculated
using the van't Hoff-Arrhenius law within the harmonic approximation
\cite{Vineyard1,Lee1}. Herein $\nu_{i}^{IS}$ and $\nu_{i}^{TS}$ are,
respectively, the normal-mode frequencies at the initial and
transition states, and can be obtained from the phonon frequency
calculation at each configuration. The calculated attempt frequency,
$\nu_{\rm G}$, of the vacancy migration ($\thicksim$
$7\times10^{13}$ s$^{-1}$) is higher than that of the carbon adatom
($\thicksim$ $3\times10^{13}$ s$^{-1}$). This is because the bonding
between the edge atom and its adjacent atoms is stronger than that
between the adatom and its adjacent atoms, leading to a higher
vibrational frequency $\nu_{i}^{IS}$ (Note: the migration of the
vacancy can be viewed as the migration of the carbon edge atom). The
calculated rate coefficient of the adatom migration ($k$ $\thicksim$
$5$ s$^{-1}$ ) is much larger than that of the vacancy migration
($k$ $\thicksim$ $2\times10^{-22}$ s$^{-1}$) at room temperature
($T=300$ K). This clearly demonstrates that the carbon adatom is
quite mobile but the vacancy is almost immobile, consistent with the
experimental observation that adatoms are generally much more mobile
in the bulk or at the surface than vacancies \cite{Banhart1}.

\begin{figure} [tbp]
\centering
\includegraphics[width=0.47\textwidth]{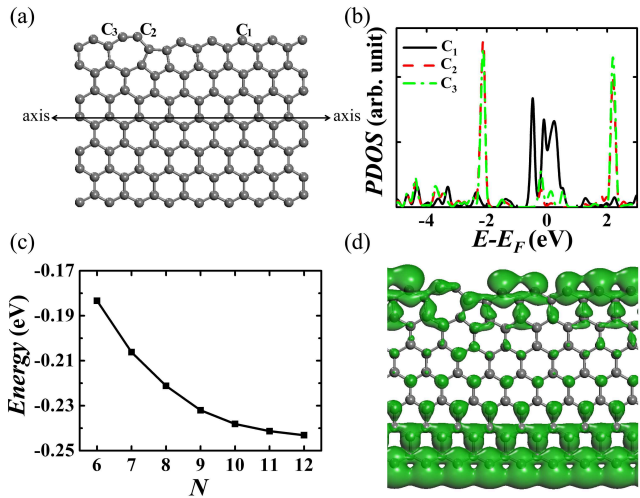}
\caption{\label{fig:fig2} (Color online) (a) The R57 which exists at
one edge of the 7-ZGNR. (b) The PDOS of three edge carbon atoms
[C$_1$, C$_2$, C$_3$, as indicated in (a)]. (c) Calculated energy
difference between the $N$-ZGNRs with one edge R57 defect and that
with perfect edges as a function of the ribbon width (i.e., index
$N$). (d) The isosurface of charge density ($0.01$ e/\AA$^3$) within
a $0.25$ eV energy window (from $-0.125$ to $0.125$ eV relative to
Fermi energy) in the 7-ZGNR with one edge R57 defect.}
\end{figure}

It should be noted that the edge adatoms and vacancies may recombine
during their migration, which results in the formation of a stable
edge structure. Without loss of generality, we consider the
recombination process of the adatom-vacancy pair where an adatom and
a vacancy are adjacent to each other. Figure~\ref{fig:fig1}(c) shows
the MEP of the recombination, as well as the initial structure (AVP)
and final structure (perfect edges). It can be seen that the 7-ZGNR
with AVP is about 1.90 eV higher in energy than that with perfect
edges. The MEP of the recombination is quite complicated, compared
with those of migrations of the adatom and vacancy. Interestingly,
it is found from the MEP that there exists a structure more stable
(about 0.21 eV lower in energy) than the 7-ZGNR with perfect edges.
Figure~\ref{fig:fig2}(a) shows the detailed structure corresponding
to the minimum total energy in the MEP, with one R57 defect at the
edge. This result is somewhat unexpected since for perfect carbon
nanotubes (CNTs) or graphene, the introduction of the Stone-Wales
defect (two pairs of R57) will always increase the total energy of
CNTs or graphene \cite{Zhou1,Crespi1}. Two very recent works have
predicted that graphene could have periodic R57 edge structure
\cite{Wassmann1,Koskinen1}, which supports our result to some
extent. However, the difference is that the R57 found here is not
periodic and is formed from the recombination of the adatom and
vacancy. Note that edge adatoms and vacancies can occur with high
probability when ZGNRs are prepared in experiments. Thus, in the
case of no hydrogenation of edges, local R57 structures are very
likely to appear in practice as the adatoms migrate along the edge
and the recombination of the adatom and vacancy proceeds.

As shown in Fig.~\ref{fig:fig1}(c), the energy barrier of the
recombination (i.e., from the AVP to R57 edge structure) is about
0.89 eV \cite{Barrier-check1}. The calculated rate coefficient $k$
is $\thicksim$ 0.1 s$^{-1}$ at room temperature, fast enough to be
observed in experimental time scale. Importantly, our further
calculations indicate that the barrier and the rate coefficient are
insensitive to the widths of ZGNRs. It is thus highly possible to
form a R57 structure from the AVP at the edges of ZGNRs. In fact,
two very recent experiments have already demonstrated that the
migration or vaporization of carbon edge atoms plays an important
role of edge reconstruction of graphene and GNRs\cite{Jia1,Girit1}.
Especially, the pentagon-heptagon ring can be clearly identified at
the zigzag edge in transmission electron microscope images of
graphene [see the supporting materials of Ref.~ \onlinecite{Girit1}
and Fig. 1 of Ref.~ \onlinecite{Koskinen}], which supports our
theoretical prediction of the stable R57 edge defect. On the other
hand, we should note that although the perfect zigzag edge is less
stable than the R57 edge structure in energy, the energy barrier of
the transformation from the perfect zigzag edge to the R57 structure
is quite high [$\thicksim$ 1.61 eV as shown in
Fig.~\ref{fig:fig1}(c)], and the corresponding rate coefficient is
small ($k\thicksim 1\times10^{-13}$ s$^{-1}$ at room temperature).
This means that once the perfect zigzag edges are formed, they are
relatively stable and will not transform to the R57 edge structure
at room temperature for a long time. Such a feature is just
consistent with the recent microscopy observations of the perfect
zigzag edges appearing under special experimental conditions
\cite{Jia1,Girit1}.

The high stability of the R57 edge, superior to the perfect edge,
can be well understood by the formation of new bonds between edge
atoms. Compared with 7-ZGNR with perfect edges, two edge atoms in
the R57 structure [i.e., $\rm{C_2}$ and $\rm{C_3}$ in Fig. 2(a)] can
form an additional bond since both of them have the dangling $sp^2$
orbitals. This can be clearly seen from two sharp peaks
(representing bonding and anti-bonding states, respectively) in the
projected density of states (PDOS) of such two atoms
[Fig.~\ref{fig:fig2}(b)]. The calculated $\rm{C_2}$$-$$\rm{C_3}$
bond length is about 1.23 \AA, typical for $-$C$\equiv$C$-$ triple
bond, which is consistent with the result of the periodic R57
structure \cite{Wassmann1,Koskinen1}. Therefore, the decrease in
energy of the R57 edge structure reflects a compromise between the
decrease in the total energy as a result of the formation of new
bonds and the increase in energy due to the structural deformation
of R57. In addition, our calculations show that the R57 structure
can also exist at the open end of zigzag CNTs. For example, the
total energy of open-ended (10,0) zigzag CNT with one R57 edge
structure is 0.15 eV lower than that with perfect edge structure.
Moreover, the energy difference between the ZGNRs with one R57 edge
structure and with perfect edges monotonously decreases with the
ribbon width [as shown in Fig. 2(c)]: the wider the ZGNR is, the
more stable the R57 edge structure is \cite{LDA}. This is because
the stress induced by structural deformation can be released more
easily as the width of ZGNRs increases. Typically, the energy
difference is about $-0.24$ eV per R57 defect for the 12-ZGNR (about
2.4 nm wide).

\begin{figure} [tbp]
\centering
\includegraphics[width=0.47\textwidth]{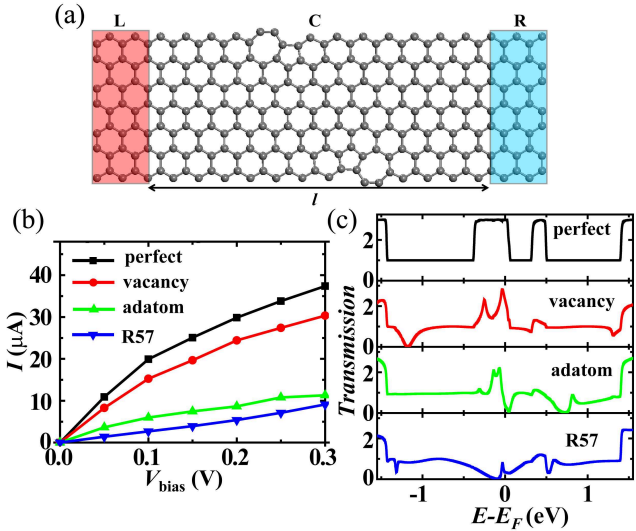}
\caption{\label{fig:fig3} (Color online) (a) Schematic structure of
a two-probe 7-ZGNR system, where edge defects exist at the both
edges. A bias voltage $V_{\rm bias}$ is applied to the central
region with the length ($l$) of 14 unit cells. (b) $I$-$V_{\rm
bias}$ curves and (c) Transmission spectra under $V_{\rm bias}$=0.1
V for 7-ZGNRs with perfect or defective edges.}
\end{figure}

For ZGNRs with perfect edges, two partially flat bands around the
Fermi level, which come from $\pi$/$\pi^{\ast}$ electrons and are
mostly localized at the two edges for large wave vector (i.e., edge
states), are responsible for the electron transmission under low
bias voltages ($V_{\text{bias}}$) \cite{Zuanyi1}. The presence of
edge defects will change the $\pi$/$\pi^{\ast}$ state and profoundly
influence the conduction mechanism \cite{Martins1}. To understand
the role of edge defects in the transport behaviors of ZGNRs under
bias voltages, we carry out extensive first-principle calculations
of $I$-$V_{\text{bias}}$ curves by using a two-probe system
[Fig.~\ref{fig:fig3}(a)], where left (L) and right (R) leads are
semi-infinite ZGNRs, and the central region (C) is a 7-ZGNR of 14
unit cell length with two edge defects (one at each edge). As shown
in Fig.~\ref{fig:fig3}(b), the current ($I$) of the ZGNR with
perfect edges increases rapidly as bias voltage increases, because
there are at least one conductance quantum $G_{0}$ (arising from the
$\pi$/$\pi^{\ast}$ state) around the Fermi level (extra 2$G_{0}$
arises from dangling bond states of edge atoms)
[Fig.~\ref{fig:fig3}(c)]. The presence of any one kind of edge
defects, however, always leads to a decrease in the current. For
example, for $V_{\text{bias}}$=0.1 V, the currents of 7-ZGNRs with
vacancy, adatom and R57 edge defects are 15.26 $\mu$A, 5.98 $\mu$A
and 2.69 $\mu$A respectively, which are all smaller than the current
of 19.93 $\mu$A for the 7-ZGNR with perfect edges. More importantly,
the current of the ribbon with the R57 defects drops drastically and
the system even deviates from original metallic transport behavior
of ZGNRs. This can be understood from the transmission spectrum
(i.e., quantum conductance) of the 7-ZGNR under a certain bias
voltage. As shown in Fig.~\ref{fig:fig3}(c), the transmission around
the Fermi level for the R57 edge defect is much smaller than that
for other cases. As mentioned before, the transport behavior of
ZGNRs under low bias voltages is determined by the
$\pi$/$\pi^{\ast}$ state related to the edge states. Among three
kinds of edge defects we discuss here, R57 edge defect destroys
original hexagonal structure along the edges most seriously, and
thus changes the original property of the $\pi$/$\pi^{\ast}$ state
notably [as shown in Fig.~\ref{fig:fig2}(d)], and causes a
remarkable conductance suppression around the Fermi level.

\begin{figure} [tbp]
\centering
\includegraphics[width=0.47\textwidth]{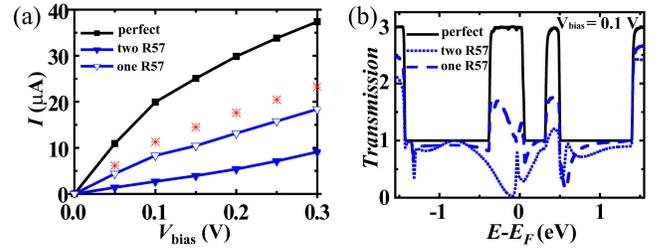}
\caption{\label{fig:fig4} (Color online) (a) $I$-$V_{\rm bias}$
curves and (b) transmission spectra under $V_{\rm bias}$=0.1 V for
7-ZNGRs with perfect edges and with two R57 defective edges, and
7-ZGNR with one R57 defective edge and one perfect edge. The red
stars in (a) indicate the average current of 7-ZGNRs with perfect
edges and with two R57 defective edges under the same bias voltage.}
\end{figure}

We also calculate the $I$-$V_{\text{bias}}$ curve and transmission
spectrum for the 7-ZGNR containing only one R57 defect. The results
are shown in comparison with those for the ribbons with two perfect
edges and with two defective edges in Fig.~\ref{fig:fig4}. Under low
bias voltages, the current of 7-ZGNR with one R57 edge defect is
between those of 7-ZGNR with perfect edges and 7-ZGNR with two R57
defects at both edges, but a little less than their average value.
This indicates that the R57 edge defect can significantly reduce the
conductance along the defective edge, but the conduction channel
along the perfect edge without R57 defect is less affected even for
narrow ribbons. The conduction characteristic can be understood from
the transmission spectra (i.e., quantum conductance) of the 7-ZGNRs
with three kinds of edges mentioned above. Under the same bias
voltage, the conductance around the Fermi level of 7-ZGNR with one
R57 edge defect is smaller than that with perfect edges, but larger
than that with two R57 defective edges. These results further verify
the validity of our conclusion that the conductance is drastically
suppressed by R57 edge defects in ZGNRs. Moreover, similar transport
behaviors are also observed for 11-ZGNRs and fully H-passivated
7-ZGNRs with different defective edges (data not shown). These
results confirm a drastic conductance suppression by R57 edge
defects in ZGNRs.

Recent experiments showed that all narrow GNRs (from sub-10 nm to
several tens nanometer) by lithographical fabrication or chemical
growth are semiconductors without evident crystallographic
directional dependence \cite{Han1,Dai,Tapaszto}, while numerous
theoretical works predicted all perfect ZGNRs are
metallic\cite{edge-state}. This implies a crucial role of the
detailed edge structure in the transport properties of GNRs. In
fact, under current experimental techniques, it is still very
difficult to obtain GNRs with atomically smooth edges. Thus, our
transport calculations based on the used model cannot lead to a
direct comparison with present experimental measurements of GNRs,
where the edge disorder is generally believed to play a key role in
the transport characteristic of GNRs. Considering its high
stability, however, we expect that the R57 edge defect may also be
an important factor to determine the practical transport behavior of
fabricated GNRs. Further studies on other kinds of edge defects as
well as edge defect complexes are still needed to fully understand
the overall effect of edge defects on the transport properties of
GNRs.

\section{Conclusions}

In summary, we have presented a detailed study on the migration of
edge defects (adatom and vacancy) and the recombination of
atom-vacancy pair in ZGNRs. The migration barriers of adatom and
vacancy are 0.76 eV and 2.12 eV, respectively. The recombination
process of the adatom-vacancy pair through the adatom migration,
which could occur at room temperature by overcoming an activation
barrier of 0.89 eV, will produce a more stable R57 edge structure.
It indicates that the ideal ZGNR is less stable than the formed
edge-defect structure. The current of ZGNRs with R57 edge defects
drops drastically as a result of the notable suppression of
conductance around the Fermi level. Our results, indicating a
possible stable edge defect structure in experimentally fabricated
ZGNRs, provide an explicit mechanism (in addition to the edge
disorder) to understand why all narrow GNRs prepared in experiments
are semiconductors.

\section*{ACKNOWLEDGMENTS}

We acknowledge the support of the Ministry of Science and Technology
of China (Grant No. 2006CB605105 and 2006CB0L0601), the National
Natural Science Foundation of China (Grant No. 10674077), and the A3
Foresight Program of KOSEF-NSFC-JSPS. J.I. was supported by the SRC
program of MEST (Center for Nanotubes and Nanostructured
Composites).


\begin{references}
\bibitem{Barrier-check1} To check the validity of our model used,
we do complementary energetics calculations for 7-ZGNR with two
defective edges. The obtained recombination barrier of an AVP pair
is almost the same as that of the 7-ZGNR with one perfect edge and
the defect being located on the other edge.
\end{references}
\end{document}